QUANTUM MAGNETISM
# Thermal Hall conductivity of $\alpha$-RuCl$_3$

Hae-Young Kee

*Department of Physics, University of Toronto, Toronto, Ontario, Canada*

**Thermal Hall conductivity originating from topological magnons is observed in the Kitaev candidate $\alpha$-RuCl$_3$ in broad intervals of temperature and in-plane magnetic field, raising questions on the role of the Majorana mode in heat conduction.**

The black-coloured ruthenium trichloride (α-RuCl$_3$) has a layered honeycomb structure composed of Ru$^{3+}$ with a magnetic moment of an effective spin-1/2. Although RuCl$_3$ compounds were discovered back in the early twentieth century, physicists only began to perceive their connection to the Kitaev spin liquid (KSL) — a special kind of quantum spin liquid (QSL) — in 2014[1]. The elementary excitations of a KSL, Majorana fermions and vortices, offer a platform for quantum memory protected from decoherence, as they cannot be annihilated locally but only through fusion with their antiparticle[2]. The smoking-gun signature of the KSL is 1/2-integer quantized thermal Hall conductivity under a magnetic field, originating from unpaired Majorana moving around the edge of the sample[2]. Remarkably, observation of 1/2-integer quantized thermal Hall conductivity in narrow ranges of temperature and magnetic field has been reported[3]. However, such experiments were repeated by a few groups using an in-plane magnetic field[4-6], and conflicting conclusions were drawn. Despite similar-looking data, one group concluded robust 1/2-integer quantization[4], while another reported no trace of 1/2-integer quantization[5], which has generated considerable debate in the community of quantum magnetism.

The thermal Hall experiment measures the temperature change ($\Delta T$) transverse to the thermal current ($\mathbf{J}_Q$) applied in the sample under a magnetic field (**B**) (Fig. 1a) which generally measures magnetic excitations in magnetic materials. When spins are ordered or partially aligned by an external magnetic field, that is, a polarized state, the low-energy excitation is a collective motion of the spins, referred to as magnon. A topological magnon is characterized by a finite Chern number associated with the Berry phase in momentum space (Fig. 1b) and may exist in the high magnetic field region of the phase diagram. This means that a magnon propagating transverse to the thermal current leads to a finite thermal Hall conductivity with a temperature dependence following bosonic statistics. To differentiate the source of heat carriers, a detailed measurement of the thermal Hall conductivity is required. Writing in *Nature Materials*, Peter Czajka and colleagues[7] report a comprehensive measurement of the thermal Hall conductivity over broad intervals in temperature and in-plane magnetic field ($B_a$) and demonstrate that the finite but not quantized thermal Hall signal arises from topological magnons, in contrast to the earlier report of the Majorana mode being the heat carrier. A theoretical study also found topological magnons in the polarized state using a widely accepted set of spin exchange parameters for α-RuCl$_3$[8], consistent with the conclusion drawn by Czajka and colleagues[7].

Czajka and colleagues further suggest that there may be a QSL in the intermediate field region bounded by critical magnetic field strengths $B_{c1}$ and $B_{c2}$ at very low temperatures (Fig. 1b), where deviation from the expected magnon occurs, and oscillation of the longitudinal thermal conductivity is observed[5]. As the transition between the polarized state and the QSL at $B_{c2}$ is only well defined at $T = 0$ K in two dimensions, this implies that there may be a crossover between a QSL and the polarized state as the temperature increases, where the topological magnons become responsible for the finite thermal Hall signal.

From microscopic theory, the intermediate field-induced KSL is unexpected, as the so-called vison gap protecting the KSL is about 0.07K, where K is the Kitaev interaction[2], implying that the KSL is fragile upon introducing other perturbations. Indeed, α-RuCl$_3$ shows a magnetic ordering with a zigzag pattern in lieu of the KSL at low temperatures[9-11], despite the dominant Kitaev interaction[1,12-15]. The survival of the KSL is even less likely when the interaction is ferromagnetic. For example, a field of about 0.02K destroys the KSL and turns it into the polarized state[16], whereas for an antiferromagnetic interaction, the KSL is extended up to a field strength of about 0.3K[17-23]. However, it is possible that non-Kitaev interactions work together with the Kitaev interaction and promote a QSL under a magnetic field. Such possibilities have been investigated by several theoretical groups using various numerical techniques[24-28]. Given the huge phase space of exchange parameters, the focus was near the ferromagnetic Kitaev interaction regime relevant for α-RuCl$_3$. A direct transition from the zigzag order to the polarized state was found when the magnetic field is applied in the plane[25,26], contradicting the experimental observations. Strikingly, when the field is oriented out of the plane, a magnetically disordered intermediate phase was found[26-28]. The strong anisotropic field response is due to the non-Kitaev interaction called Γ[26,28,29]. Whether the intermediate state boosted by the positive Γ interaction is a QSL or not remains to be resolved, as a finite Γ allows for mobile visons, and the free Majorana picture of the pure Kitaev model does not work. However, the effects of the Γ interaction and a magnetic field somehow cancel, and the field-induced intermediate phase may map to the effective KSL with a perturbing magnetic field. While this scenario seems unlikely, it has not been ruled out.

There are experimental challenges owing to a strong sample dependence[30,31]. The layers of α-RuCl$_3$ are stacked via a weak van der Waals interaction and different types of stacking are naturally expected[10,11,30-34]. Depending on the stacking pattern of α-RuCl$_3$, the in-plane spin exchange parameters vary, because of the changes in the Ru–Ru ion bond length and the angle between Ru–Cl–Ru bonds[32]. As this sensitivity traces back to the spin–orbit entangled wavefunction[35], it is difficult to avoid. If the Kitaev interaction is antiferromagnetic in certain samples, a more robust spin liquid and a proposed U(1) spin liquid may occur under the magnetic field[21-23]. If so, the moment direction in the ordered states may differ from the samples with a dominant ferromagnetic Kitaev interaction[36,37]. Looking ahead, thorough experimental studies on a given sample that give a full set of information, including the layer stackings, the moment direction of the magnetic order, the anisotropy in the susceptibilities, the dynamic excitations and the thermal Hall measurements in different field directions, will advance our search for a material realization of a KSL.

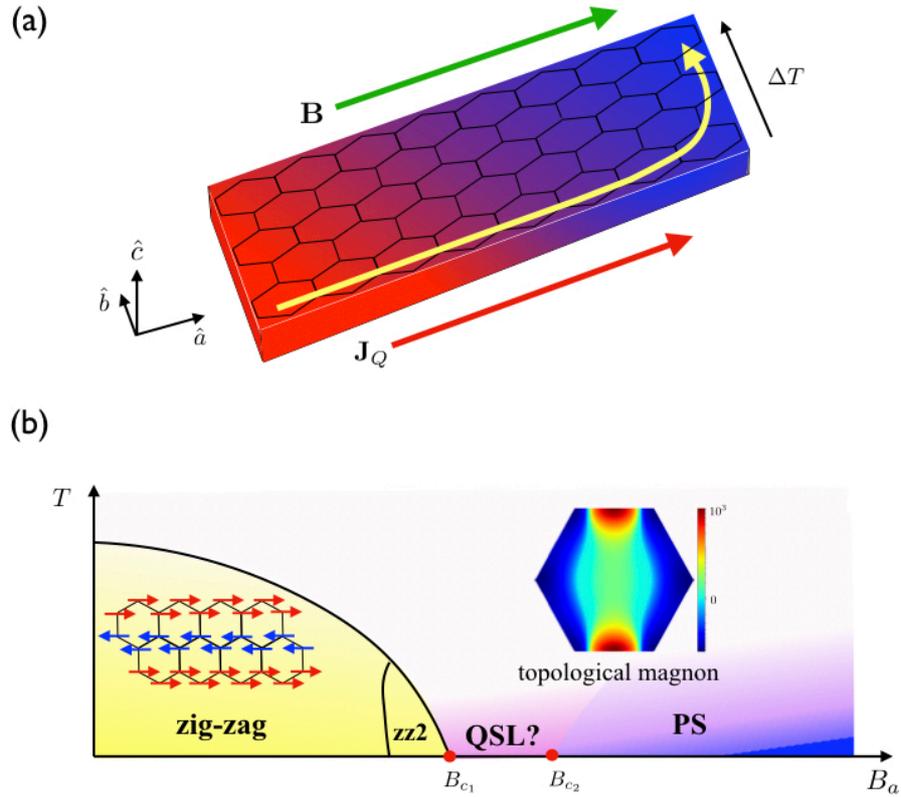

**Figure 1 Thermal Measurements and Phase Diagram of α-RuCl3**. **a**, Thermal Hall experimental set-up. The magnetic field (**B**) and thermal current (**J**$_Q$) are applied along the a axis (in-plane direction perpendicular to one of the honeycomb bonds), and the change of temperature ($\Delta T$) is measured across the b axis. The yellow arrow represents a heat carrier leading to a finite thermal Hall conductivity. For the other in-plane direction, that is, the b axis (parallel to one of the honeycomb bonds) set-up, the thermal Hall conductivity vanishes due to the ac-mirror plane (or $C_2$ rotation about the b axis) symmetry. **b**, Phase diagram with respect to the a-axis field strength ($B_a$) and temperature ($T$). At low temperatures without the magnetic field, a magnetic order with a zigzag pattern is found, indicated by red and blue arrows. As the field strength increases, the zigzag pattern along the layers is modified (zz2)[34], indicating a weak interlayer spin interaction altered by the in-plane field. A puzzling QSL between $B_{c1}$ and $B_{c2}$ before the polarized state (PS) was suggested at very low temperatures. As temperature increases, a crossover to the polarized state may occur. The rainbow-coloured hexagon in the polarized state denotes the Berry curvature in the first Brillouin zone, leading to a finite thermal Hall conductivity.